\documentclass[a4paper]{article}

\usepackage[english]{babel}
\usepackage[utf8x]{inputenc}
\usepackage[T1]{fontenc}
\usepackage{authblk}
\usepackage{parskip}
\usepackage[normalem]{ulem}

\usepackage[a4paper,top=3cm,bottom=2cm,left=3cm,right=3cm,marginparwidth=1.75cm]{geometry}
\usepackage[onehalfspacing]{setspace}

\usepackage{amsmath}
\usepackage{mathtools}
\usepackage{graphicx}
\usepackage[colorinlistoftodos]{todonotes}
\usepackage[colorlinks=true, allcolors=blue]{hyperref}
\usepackage[authoryear,comma,semicolon]{natbib}
\usepackage{dsfont}

\usepackage{caption} 
\usepackage{subcaption}

\title{A flexible adaptive lasso Cox frailty model based on the full likelihood }

\author[1]{Maike Hohberg\thanks{mhohber@uni-goettingen.de}}
\author[2]{Andreas Groll\thanks{groll@statistik.tu-dortmund.de 
\vspace{12pt}\newline
}
}

\affil[1]{Chair of Statistics, University of Goettingen}
\affil[2]{Statistical Methods for Big Data, TU Dortmund University}

\begin{document}
\maketitle

\begin{abstract} \noindent
In this work a method to regularize Cox frailty models is proposed that accommodates time-varying covariates and time-varying coefficients and is based on the full instead of the partial likelihood. A particular advantage in this framework is that the baseline hazard can be explicitly modeled in a smooth, semi-parametric way, e.g.\ via P-splines. Regularization for variable selection is performed via a lasso penalty and via group lasso for categorical variables while a second penalty regularizes wiggliness of smooth estimates of time-varying coefficients and the baseline hazard.   
Additionally, adaptive weights are included to stabilize the estimation.
The method is implemented in \texttt{R} as \texttt{coxlasso} and will be compared to other packages for regularized Cox regression. Existing packages, however, do not allow for the combination of different effects that are accommodated in \texttt{coxlasso}.        
\end{abstract}

%%%%%%%%%%%%%%%%%%%%%%%%%%%%%%%%%%%%%%%%%%%%%%%%%%%%%%%%%%%%%%%%%%%%%%%%%%%%%%%%%%%%%%
%%%%%%%%%%%%%%%%%%%%%%%%%%%%%%%%%%%%%%%%%%%%%%%%%%%%%%%%%%%%%%%%%%%%%%%%%%%%%%%%%%%%%%

\section{Introduction} 

Cox's well-known proportional hazards model \citep{cox.1972} assumes the semi-parametric hazard
\begin{align} \label{eq:cox1}
\lambda(t|\mathbf{x}_i) = \lambda_0(t) \exp(\mathbf{x}_i^{\top} \boldsymbol \beta)\,,
\end{align}
where $\lambda(t|\mathbf{x}_i)$ denotes the hazard function for individual $i$ at time $t$ conditional on covariates $\mathbf{x}_i$. The shared baseline hazard
$\lambda_0(t)$ is usually not further specified 
and $\boldsymbol \beta$ is a vector for $p$ fixed effects.
%The hazard rate $\lambda(t|\mathbf{x}_i)$ can be written as the instantaneous risk of transition at time $t$  i.e.
%\begin{align}
%\lambda(t|\mathbf{x}_i) = \lim_{\Delta t \to 0} P(t\leq T < t + \Delta t | T \geq t, \mathbf{x}_i) / \Delta t
%\end{align}
%in the case of continuous time. 
Estimation of the model is typically based on maximizing the partial likelihood which has the advantage of removing  $\lambda_0(t)$ from the estimation of $\boldsymbol \beta$. For the case of a large number of predictors, the lasso penalty was incorporated into the Cox model to enable variable selection and shrinkage \citep{Tibshirani.1997}.  Different algorithms to fit the penalized model have been proposed, for example, by \citet{Gui.2005} using  least-angle regression (LARS), by \citet{Simon.2011} via a coordinate descent algorithm, or by \citet{Goeman.2010} who combines gradient ascent optimization with the Newton-Raphson algorithm.  

These exiting penalization procedures are all based on the partial likelihood. In samples with a small or moderate number of observations, using the partial likelihood can lead to loss in efficiency and precision  samples \citep[see, e.g.,][]{cox.1984,Ren.2011}. Considering the popularity of regularized Cox models and given that in many medical applications the sample size is often rather small, it seems surprising that there is, to the best of our knowledge, currently no available implementation that uses a simple lasso penalty within the Cox full likelihood model. Note that strictly spoken, the term Cox model inherently implies the use of the partial likelihood and using the full likelihood would thus correspond to a different model. However, for convenience we still use the term Cox model even in the context of the full likelihood.  %Especially for datasets with a small or moderate number of observations, using the full likelihood does not drastically increase computing time.

Despite the predominance of the partial likelihood in existing \texttt{R} routines, 
there are some advantages when using the full likelihood:
    \begin{enumerate}
    \item the baseline hazard can be modeled explicitly, e.g., using a basis function approach such as B-splines \citep[see e.g.][]{Eilers.1996},
    \item the full likelihood model can easily be extended by a wide class of frailty distributions including random intercepts and random slopes, 
    \item (time-)varying covariates can be naturally incorporated. 
    \end{enumerate}

%The purpose of this paper is thus twofold: We first show that in the case of changing covariates using the full likelihood is preferable. This effect increases with the frequency of the covariates' changes. Secondly, we implement the full likelihood in a package \texttt{coxlasso} that includes a classical lasso regularization and can easily accommodate changing covariates, frailties, and time varying coefficients.  

In a simulation study, in which we analyze the validity of the approach,  we also assess  %instead of focusing on performance in small and moderate samples, we focus on another 
a condition that has not attracted much attention yet and in which the full likelihood might be beneficial: We examine the situation where we have covariates that change their status relatively frequently, which potentially affects the survival outcome. The reason why the full likelihood becomes relevant when covariates change frequently becomes apparent when comparing partial and full likelihood. In survival analysis, we typically deal with data consisting of a tuple $(T_i, d_i)$ with $d_i$ indicating whether an event happened, i.e.\ $d_i = 1$ if the survival time is completely observed, whereas $d_i = 0$ if this observation is right-censored. $T_i$ is a random variable and can be described by event time $\tilde T_i$ and censoring time $C_i$ via $T_i=\min(\tilde T_i,C_i)$ and $d_i = \mathds{1}(\tilde T_i \leq C_i)$.
%, with $T_i=\min(\tilde T_i,C_i)$.
For each tuple  $(T_i, d_i)$, the density $f(T_i,d_i)$ can be written as   
\begin{align*}
f(T_i,d_i) & =\left(f(\tilde T_i) P(\tilde T_i<C_i)\right)^{d_i} \cdot \left(g(C_i) P(\tilde T_i \geq C_i) \right) ^{1-d_i} \propto f(\tilde{T}_i)^{d_i} \cdot S(\tilde{T}_i)^{(1-d_i)} \,\\
&= \left(\lambda_0(\tilde{T}_i) \exp (\mathbf{x}_i^{\top} \boldsymbol \beta)\right)^{d_i} \, \exp\left(- \int_0^{\tilde{T}_i} \lambda_0(s) \exp(\mathbf{x}_i^{\top} \boldsymbol \beta) ds\right)\,, 
\end{align*}
with $f(t) = \lambda(t)S(t)$, where $S(t)$ is the survivor function. 
The full likelihood over all individuals thus yields
\begin{align} \label{eq:fulllik}
L(\lambda_0(t), \boldsymbol \beta) &= \prod_{i=1}^n \left(\lambda_0(t_i) \exp (\mathbf{x}_i^{\top} \boldsymbol \beta)\right)^{d_i} 
\underbrace{\exp \left(-\int_0^{t_i} \lambda_0(s) \right) ds}_{:=S_0(t_i)} \exp(\mathbf{x}_i^{\top} \boldsymbol \beta)\nonumber\\
&=  \prod_{i=1}^k \frac{\exp(\mathbf{x}_i^{\top} \boldsymbol \beta)}{\sum_{j \in R(t_{(i)})}\exp(\mathbf{x}_{j}^{\top}\boldsymbol \beta)}
\cdot \left(\sum_{j \in R(t_{(i)})} \lambda_0(t_{(i)}) \exp(\mathbf{x} _{j}^{\top}\boldsymbol \beta)
\cdot \prod_{i=1}^n S_0(t_i) \exp (\mathbf{x}_i^{\top} \boldsymbol \beta) \right)\,,
\end{align}
where $t_i, i=1,\ldots,n,$ are the observed censoring or event times of each individual, $t_{(1)}< \dots <  t_{(i)} < \dots <  t_{(k)}$ denote the ordered distinct event times of the uncensored individuals, and $R(t)$ is the set of individuals under risk at time $t$. The first factor of equation (\ref{eq:fulllik}), which is the ratio of the event probabilities of all individuals that died at time $t$ and the event probabilities of all individuals at risk at that time, corresponds to the {\it partial likelihood}. Besides removing the baseline hazard from the inference process of $\boldsymbol \beta$, the partial likelihood is attractive since covariate information can be easily included and it is not affected by the censoring pattern \citep{Efron.1977}.  However, it actually is not a real likelihood as it  ignores the integral part of the full likelihood and, hence, certain covariate information from non-failure intervals and is thus not based on all observations. Since the partial likelihood nearly contains all of the information about $\boldsymbol\beta$, the estimate $\hat{\boldsymbol\beta}$ is still asymptotically efficient. 

However, relying on the partial likelihood ignores the second factor of equation~\eqref{eq:fulllik}.
So if there is a lot of information in non-failure intervals and if these might influence the survival outcome, the partial likelihood might not give satisfying estimates. In particular, time-varying covariates result in splits of the data and could create several new, censored observations.
The more often covariates change, the more splits get neglected since the partial likelihood only considers the status of the covariates at the event times but not in between.  

The literature on the full likelihood for regularized Cox models is rather scarce while a few approaches exist for the unregularized Cox model. For example, \citet{Ren.2011} proposed a maximum likelihood (ML) estimator based on the full-profile likelihood function for $\boldsymbol \beta$ which they obtained by profiling out the nuisance parameter $F_0$, the baseline distribution associated with $\lambda_0$. \citet{Gentleman.1991} propose a local full likelihood method alternating between estimating the baseline hazard and the covariate effects. Other approaches that use the full likelihood to accommodate a baseline hazard modelled via splines include \citet{Etezadi1987}, \citet{Rosenberg1995}, \citet{Herndon1995}, and \citet{Devarajan2011}. However, none of these analyzes the case of changing covariates nor do they provide a readily implemented \texttt{R} package. For this reason, alongside with the \texttt{coxlasso} function we also provide the function \texttt{coxFL}, which implements the (unregularized)  Cox full likelihood approach and allows for changing covariates, frailties, and time-varying coefficients. Our main contribution, however, is implemented in an \texttt{R}-function called \texttt{coxlasso} that includes a classical (adaptive) lasso penalization and can easily accommodate changing covariates, frailties, and time-varying coefficients. Currently, a working version is available directly from the authors upon request, which will soon be incorporated in the \texttt{R}-package \texttt{PenCoxFrail} (Groll, 2016). We apply \texttt{coxlasso} to two illustration examples: We first examine the determinants of exclusive breastfeeding duration in Indonesia, and the second example performs variable selection among genetic and clinical data to explain survival duration of lung cancer patients. 

%\textcolor{red}{include fan , li und Androulakis, sy and taylor for coxph cure models, EM algorithm, staniswalis: kernel estimator 
%Oder wird eigentlich immer full likelihood genommen wenn hazard mit bsplines approximiert wird? bsp angelos 1991, rosenberg 1995}

The remainder of this manuscript is structured as follows: Section 2 presents the model and the estimation procedure. Simulations are carried out in Section 3 while Section 4 shows an application of \texttt{coxlasso}. Finally, Section 5 concludes.

%%%%%%%%%%%%%%%%%%%%%%%%%%%%%%%%%%%%%%%%%%%%%%%%%%%%%%%%%%%%%%%%%%%%%%%%%%%%%%%%%%%%%%
%%%%%%%%%%%%%%%%%%%%%%%%%%%%%%%%%%%%%%%%%%%%%%%%%%%%%%%%%%%%%%%%%%%%%%%%%%%%%%%%%%%%%%

\section{Methodology} 
  
In the following, we sketch the model and explain how different effects are included. We also elaborate on the applied estimation algorithm based on the full likelihood.    
  
\subsection{Model} 

We specify a flexible model for the conditional hazard function as follows:  
\begin{align} \label{eq:coxfrailvary}
 \lambda(t| \mathbf{x}_{ij}, \mathbf{z}_{ij}, \mathbf{u}_{ij}, \mathbf{b}_{i})
&= \exp(\eta_{ij}(t))\,, 
\qquad i = 1, \dots, n; j = 1, \dots, N_i
\end{align}
with corresponding predictor for individual $j$ belonging to cluster $i$
\begin{align} \label{eq:pred}
\eta_{ij}(t) &=  \gamma_0(t) + \mathbf{x}_{ij}^{\top} \boldsymbol \beta +  \mathbf{z}_{ij}^{\top}\boldsymbol \gamma(t)  + \mathbf{u}_{ij}^{\top} \mathbf{b}_i\,,
\end{align}
%
%\sum_{k=1}^K {v}_{ijk}\gamma_k(t)
%
where $\gamma_0(t)$ is the (logarithmized) baseline hazard, $\mathbf{x}_{ij}^{\top} \boldsymbol \beta$ represent linear effects, $ \mathbf{z}_{ij}^{\top}\boldsymbol\gamma(t)$ stands for covariate effects varying over time and $\mathbf{u}_{ij}^{\top} \mathbf{b}_i$ and represent random effects. 

%Let now $\boldsymbol\alpha^{{\top}}=(\boldsymbol\alpha_0^{{\top}},\boldsymbol\alpha_1^{{\top}},\ldots, \boldsymbol\alpha_r^{{\top}})$
%collect all spline coefficients corresponding to baseline hazard and time-varying effects in case that such are included. 

Analogously to equation~\eqref{eq:fulllik}, the full likelihood for model (\ref{eq:coxfrailvary}) is given for a single cluster $i$ by
\begin{align} \label{eq:likhood}
L_i(\boldsymbol \vartheta) = \prod_{j=1}^{N_i} \exp(\eta_{ij}(t_{ij})) ^{d_{ij}} \exp \left(-\int_0^{t_{ij}} \exp(\eta_{ij} (s)) ds\right)\,.
\end{align}
The vector $\boldsymbol \vartheta$ collects all parameters that are to be estimated. %  \mage{encompassing $\boldsymbol \vartheta = (\gamma_0, \boldsymbol\beta^{\top}, \boldsymbol \gamma^{\top}, \boldsymbol\delta^{\top}, \mathbf {b}_j^{\top})^{\top}$} for model \eqref{eq:coxfrailvary}. 
%\red{[Der $\boldsymbol \vartheta$ Vektor ist an dieser Stelle noch etwas problematisch, da $\gamma_0$ und $\boldsymbol\gamma$ hier eigtl.\ noch keine Parameter, sondern Funktionen von der Zeit sind; evtl.\ einfach den Part in magenta rausl\"oschen; die genaue definition von $\boldsymbol \vartheta$ kommt dann halt erst weiter unten...]}

The corresponding log-likelihood can be maximized using a penalized quasi-likelihood approach proposed by \citet{Breslow.1993}. 
%, that involves the  marginal log-likelihood given by
%
%\begin{align}
%l^{mar}(\boldsymbol\delta, \boldsymbol\theta) = \sum _{j=1}^N \log\left(\int L_j(\boldsymbol\beta, \boldsymbol\alpha, \mathbf{b}_j) p(\mathbf{b}_j|\boldsymbol \theta)d\mathbf{b}_j\right),
%\end{align}
%depending on parameter vector $\boldsymbol \delta^{\top} = (\boldsymbol\beta^{\top}, \boldsymbol\alpha^{\top},  \mathbf{b}^{\top})$ and on $\boldsymbol \theta$, the parameters of the covariance structure of random effects $\mathbf{b}_j$ as specified before. The density of the random effects is given by $p(\mathbf{b}_j|\boldsymbol \theta)$. Following Breslow and Clayton (1993) and
Applying Laplace approximation leads to a penalty term $\mathbf{b}^{\top}\mathbf{Q}^{-1}(\boldsymbol\theta) \mathbf{b}$ that is deducted from the likelihood contribution of each cluster, yielding an approximated log-likelihood given by
\begin{align}\label{eq:lik_app}
l^{app}(\boldsymbol \vartheta, \boldsymbol \theta) & =  \sum_{i=1}^n \log L_j (\boldsymbol \vartheta) - \frac{1}{2} \mathbf{b}^{\top} \mathbf{Q}^{-1}(\boldsymbol\theta) \mathbf{b}\,.
\end{align}
%\textcolor{red}{Frage:  In our model, the spline effects, treated as random effects, and the random cluster, or frailty effects are two different type of random components, and the marginal likelihood resulting after integrating them out does not have an analytic form. To overcome this problem Laplace approximation seems a natural candidate. However, since Laplace approximation leads to underestimation of the random effects√¢¬Ä¬ô variance, a more elaborated fitting routine is necessary. (Kauermann)}. 

In the following, we explain modeling and notation of all effect types separately  and will show how the components affect the likelihood. 

\subsubsection*{Smooth Baseline Hazard} 

We model the baseline hazard as a smooth function using B-splines. Firstly, the baseline hazard is shifted into the predictor $\eta_{ij}(t)$ using a log$(\cdot)$ transformation, $\gamma_0(t) := \log(\lambda_0(t))$, where $\lambda_0(t)$ is the untransformed baseline hazard as it appears in the simple formulation of a Cox model, i.e.\ $\lambda_{ij}(t) = \lambda_0(t) \exp(\mathbf{x}_{i,j}^{\top} \boldsymbol \beta)$. The transformed baseline hazard $\gamma_0(t)$ is then expanded in (penalized) B-splines following \citet{Eilers.1996}: 
\begin{align*}
\gamma_0(t) = \sum_{m=1}^M \alpha_{0,m}B_m(t,d)\,, 
\end{align*}
with $\alpha_{0,m} = 1, \dots, M$ denoting unknown spline coefficients associated with the $m$-th B-spline basis function $B_m(t,d)$ of degree $d$.  

\subsubsection*{Time-varying coefficients}

In the same way as we model the smooth baseline hazard, penalized B-splines can be used to represent time-varying covariate effects  $\boldsymbol \gamma(t) = (\gamma_1(t), \dots, \gamma_k(t), \dots, \gamma_K(t))^{\top}$. That is, the $k$-th time-varying effect can be expanded to 
\begin{align}\label{eq:vary:effect}
\gamma_k(t)= \sum_{m=1}^M \alpha_{k,m} B_m(t,d) \,,
\end{align}
where $k = 1, \dots, K$ indexes the number of time-varying coefficients. 

Let now $\boldsymbol\alpha=(\boldsymbol\alpha_0^{{\top}},\boldsymbol\alpha_1^{{\top}},\ldots, \boldsymbol\alpha_r^{{\top}})^{\top}$
collect all spline coefficients corresponding to baseline hazard and time-varying effects. These spline coefficients $\boldsymbol\alpha$ are included into the parameter vector $\boldsymbol\vartheta$ such that it consists now of  $\boldsymbol \vartheta = (\boldsymbol \alpha^{\top}, \boldsymbol\beta^{\top}, \boldsymbol\delta^{\top}, \mathbf {b}_j^{\top})^{\top}$. 
Additionally, to control the roughness of the smooth functions, second order differences of adjacent spline coefficients $\alpha_{k,m}$ are penalized in $J_{\zeta}(\boldsymbol\alpha)$ . The likelihood thus becomes penalized and can be written as 
\begin{align} 
l^{pen} (\boldsymbol\vartheta, \boldsymbol\theta) &= l^{app} (\boldsymbol\vartheta, \boldsymbol\theta)  -  J_{\zeta}(\boldsymbol\alpha)\,, 
\end{align}
where $J_{\zeta}(\boldsymbol\alpha) =  \zeta \cdot \boldsymbol \alpha^{\top} \boldsymbol {D}  \boldsymbol{D}^{\top} \boldsymbol{\alpha}$. The  matrix $\boldsymbol D$ is a second order difference matrix and $\zeta$ is a smoothing parameter.  

% D ist diagonal matrix

%With a ridge-type penalty, the second penalty term is thus given by
%\begin{align*}
%J_{\alpha} (\boldsymbol\alpha)  = \left(\sum_{k=0}^r  \zeta_k \,||\boldsymbol\Delta_M^2 \boldsymbol\alpha_k ||_2^2 \right), 
%\end{align*}
%with $k = 0, \dots, r$ corresponding to the baseline hazard or the  $k$-th time-varying effect. $\boldsymbol\Delta_M$ is the $((M-1) \times M)$ difference operator matrix of degree one.
%
%The tuning parameters $\zeta_k$ for the baseline hazard and time-varying coefficients, which control the smoothness of these, need to be properly chosen. One strategy would be to
%employ (multi-dimensional) cross-validation techniques, though being rather time-consuming.
%Alternatively, we 
To determine the optimal amount of smoothing, we suggest a mixed model representation of the penalized spline approach allowing data driven, fast smoothing parameter selection \citep[see, e.g.,][]{Ruppert.2003}. In this view, the regression spline coefficients $\boldsymbol \alpha_k$ that are subject to penalization are taken to be random with corresponding random effect distributions $N_k\sim(\mathbf{0}, \sigma^2_{\boldsymbol\alpha_{k}} \mathbf{I})$. The reciprocal of $\hat\sigma^2_{\boldsymbol\alpha_{k}}$ can then be used as the optimal smoothing parameter.

\subsubsection*{Frailties}

The index $j$ represents different clusters the data is grouped into, resulting in frailty component $\mathbf{b}_j$ for that particular cluster. Due to  its mathematical convenience, these frailties are often assumed  
to follow a gamma distribution, but to allow for a more flexible  predictor structure, assuming log-normally distributed frailties is more appropriate. 
Hence, we specify $\log(\mathbf{b}_i) \sim N(\boldsymbol 0, \mathbf{Q}(\boldsymbol \theta))$ with mean vector $\boldsymbol 0$ and covariance matrix $\mathbf{Q}(\boldsymbol \theta)$, where $\boldsymbol\theta$ are unknown variance-covariance parameters. The log-transformation is needed to shift the frailties into the predictor. The vector $\mathbf{u}_{ij}$ in equation~\eqref{eq:pred} contains covariates associated with frailties $\mathbf{b}_i$.

\subsubsection*{Time-varying covariates}

The covariates $\mathbf{x}_{ij}$ in equation~\eqref{eq:pred} do not need to be constant over the whole time period $[0, t_{ij}]$ but are allowed to change at several time points. Using the full likelihood has the advantage that changing covariates can be naturally incorporated.  
Including time-varying covariates will split the integral in \eqref{eq:likhood} into several sub-integrals in which $\mathbf{x}_{ij}$ are then piece-wise constant.

\subsubsection*{Lasso penalty on metric and categorical covariates} 

In order to perform variable selection and shrinkage,  a lasso-type penalty is applied to linear effects while a second penalty 
controls the wiggliness of the smooth baseline hazard (and of additional time-varying coefficients, if present). Including the penalties, the  likelihood can be written as
\begin{align} \label{eq:pen_lik}
l^{pen} (\boldsymbol\delta, \boldsymbol\theta) &= l^{app} (\boldsymbol\delta, \boldsymbol\theta)  - J_{\beta}(\boldsymbol\beta) - J_{\zeta}(\boldsymbol\alpha), 
\end{align}
where $J_{\beta}(\boldsymbol\beta) = \xi\cdot \sum_{k=1}^p w_k \,|\beta_k|$
is a lasso penalty that shrinks less important (time-constant) fixed effects $\beta_k, k= 1, \dots, p,$ towards zero and is able to exclude them from the predictor. Furthermore, $\xi \geq 0$ is a tuning parameter controlling the strength of the penalization that needs to be chosen by an appropriate technique, e.g., K-fold cross-validation (CV). Additionally, we incorporate  adaptive weights $w_k:=1/\, | \hat{\beta_k}^{\small(ML)}|$ given by the inverse of the corresponding (unpenalized) maximum likelihood (ML) estimator (if it exists; if it does not exist, a slightly ridge-penalized estimate of $\beta_k^{\small(ML)}$ can be used instead). 

If categorical predictors are present, the classical lasso penalty can be combined with a group lasso penalty \citep[see][]{Meier.2008}. In this case, the categorical variable is dummy encoded forming a group of dummies and $\boldsymbol \beta_k$ collects the corresponding coefficients of the particular group. Then, the $L_2$ norm of vector $\boldsymbol \beta_k$ is penalized yielding penalty
\begin{align}\label{eq:group:lasso}
J_{\beta}(\boldsymbol \beta) = \xi \sum_{k=1}^p  w_k \,\, \sqrt[]{df_k} \, ||\boldsymbol \beta_k ||_2 \,, 
\end{align}
where $df_k$ is the number of dummies of group $k$ and is used to rescale the penalty according to the dimensionality of $\boldsymbol \beta_k$. In this case the corresponding weights have the general form $w_k:=1/ || \hat{\boldsymbol \beta_k}^{\small (ML)}||_2$.
Of course, for a mixture of metric and categorical predictors, the conventional lasso penalty from above can also be combined with the group lasso penalty from~\eqref{eq:group:lasso}.

Note that \texttt{coxlasso} provides a solution to the classical lasso variable selection problem based on the full likelihood and allows to flexibly include a range of other effects. One of the settings relevant for \texttt{coxlasso} are applications, where we have some covariates of high interest that potentially have  time-varying effects that should be part of the model and a lot of other controls that can enter the model linearly, but are subject to variable selection. For example, in a medical setting, one can include a clinical variable of high interest with a time-varying effect and then a lot of genetic covariates of which just a few might be relevant for the model. In other settings the selection problem is designed differently. For example,   \citet{Leng.2009} and \citet{Hu.2012}  provide solutions for differentiating between constant and time-varying effects, while \citet{Tang.2012} and \citet{groll.2017}  perform variable selection and simultaneously decide whether an effect is time-varying or not. 

%%%%%%%%%%%%%%%%%%%%%%%%%%%%%%%%%%%%%%%%%%%%%%%%%%%%%%%%%%%%%%%%%%%%%%%%%%%%%%%%%%%%%%
%%%%%%%%%%%%%%%%%%%%%%%%%%%%%%%%%%%%%%%%%%%%%%%%%%%%%%%%%%%%%%%%%%%%%%%%%%%%%%%%%%%%%%

\subsection{Estimation algorithm} \label{sec:estimation_algo}

The maximization of the penalized likelihood in \eqref{eq:pen_lik} is based on a Newton-Raphson algorithm and makes use of local quadratic approximations of the penalty terms following \citet{Oelker.2017}. That is, for the classical lasso penalty one can use $|\beta_j| \approx \sqrt{\beta_j^2 +c}$, where $c$ is a very small positive number, such that the approximation is very close to the $L_1$ norm but differentiable in zero. In a similar way, also the group lasso penalty terms based on the L2 norm can be approximated.

The penalized log-likelihood from~\eqref{eq:pen_lik} with plugged-in approximated likelihood and penalties is 
\begin{align*}
\ell^{pen}(\boldsymbol \delta, \boldsymbol \theta) & =  
\sum_{i=1}^n \sum_{j=1}^{N_i} d_{ij} \eta_{ij}(t_{ij}) -\int_0^{t_{ij}} \exp(\eta_{ij} (s)) ds
- \frac{1}{2} \mathbf{b}^{\top} \mathbf{Q}^{-1}(\boldsymbol\theta) \mathbf{b} - 
\zeta \cdot \boldsymbol \alpha^{\top} \boldsymbol {D}  \boldsymbol{D}^{\top} \boldsymbol{\alpha}
- \xi \cdot  \sum_{k=1}^p   w_k   |\beta_k |\,.
%J_{\beta}(\boldsymbol\beta) - J_{\zeta}(\boldsymbol\alpha),
\end{align*}
Taking its derivative, yields the penalized score function $\boldsymbol s (\boldsymbol \delta) = \partial \ell^{pen} (\boldsymbol \delta) / \partial \boldsymbol \delta$. The vector components of $\boldsymbol s (\boldsymbol \delta)$ are: 
\begin{align*}
\boldsymbol{s_{\beta}}(\boldsymbol{\delta}) &= 
 \sum_{i=1}^n \sum_{j=1}^{N_i} \mathbf{x}_{ij}\left(d_{ij} - \int_0^{t_{ij}} \exp(\eta_{ij}(s)ds\right)  - \boldsymbol{A} \boldsymbol{\beta}, 
\\
\boldsymbol{s_{\alpha}}(\boldsymbol{\delta}) &= 
\sum_{i=1}^n \sum_{j=1}^{N_i} \left( d_{ij} \boldsymbol{\Phi}(t_{ij}) 
- \int_0^{t_{ij}} \exp(\eta_{ij}(s)) \boldsymbol{\Phi}(s) ds \right) - \boldsymbol{A}_{\zeta} \boldsymbol{\alpha}, \\
\boldsymbol{s_{b_i}}(\boldsymbol{\delta}) &=
\sum_{j=1}^{N_i} \boldsymbol{u}_{ij} \left(d_{ij}   - \int_0^{t_{ij}} \exp(\eta_{ij}(s)) ds \right)
- \boldsymbol{Q}^{-1}(\boldsymbol{\theta}) \boldsymbol{b}_i,
\end{align*}
where $\boldsymbol{\Phi}^{\top}(t):= (z_{ij0} \cdot \boldsymbol{B}^{\top}(t), z_{ij1} \cdot \boldsymbol{B}^{\top}(t), \ldots, z_{ijr} \cdot \boldsymbol{B}^{\top}(t))$ with $\boldsymbol{B}(t)$ being the design vector for the B-spline expanded on time $t$. 

The penalty matrix for the lasso penalty $\boldsymbol{A}$ is a diagonal matrix resulting from the applied approximation based on  \citet{Oelker.2017} as described in the beginning of Section \ref{sec:estimation_algo}. The matrix is built of different blocks with one block associated with all the metric covariates and then each additional categorical variable adds another block to $\boldsymbol{A}$. For metric covariates, the block is a diagonal matrix with $A_{jj}^{met} \approx \xi \cdot w_j ({\beta_j^2 +c})^{-1/2}$ on its diagonal elements.
A block for each categorical covariate has diagonal elements $A_{jj}^{cat} \approx \xi \cdot w_j \sqrt{df_j} \left(\sum_{l=1}^{df_j}\beta_{jl}^2 +c\right)^{-1/2}$. 

% additional term sqrt(levels.vec[k]) not from Oelker & Tutz, but needed to adjust group lasso to normal lasso terms (wurde im Hauptteil schon erkl√§rt)
 
The penalty matrix $\boldsymbol{A}_{\zeta}$ is a block diagonal matrix that consists of only one block if no varying coefficients are included. Then, $\boldsymbol{A}_{\zeta}$ corresponds to $\boldsymbol{A}_{\zeta_0} = \zeta_0 \cdot \boldsymbol D  \boldsymbol D^{\top}$ and represents the penalized squared differences of adjacent spline coefficients for the baseline hazard. Matrix $\boldsymbol D$ is again a second order difference matrix. If $m$ time-varying coefficients are added to the model, the penalty matrix  $\boldsymbol{A}_{\zeta}$ consists -- in addition to the block for the baseline hazard -- of blocks  $\boldsymbol{A}_{\zeta_1}, \ldots,  \boldsymbol{A}_{\zeta_m}$ that are of the same functional form as  $\boldsymbol{A}_{\zeta_0}$.     

The components of the penalized information matrix are:  
\begin{center}
$
\begin{alignedat}{2}
&\boldsymbol{F}_{\boldsymbol \beta \boldsymbol \beta} && = 
- \left(\sum_{i=1}^n \sum_{j=1}^{N_i} \mathbf x_{ij} \mathbf x^{\top}_{ij} \int_0^{t_{ij}}\exp(n_{ij}(s))ds + \boldsymbol A \right),\\
&\boldsymbol{F}_{\boldsymbol \beta \boldsymbol \alpha}  = \boldsymbol{F}^{\top}_{\boldsymbol \alpha \boldsymbol \beta} && = - \sum_{i=1}^n \sum_{j=1}^{N_i} \mathbf x_{ij}  \int_0^{t_{ij}}\exp(n_{ij}(s)) \boldsymbol \Phi^{\top} (s) ds,\\
&\boldsymbol{F}_{\boldsymbol \alpha \boldsymbol \alpha} && = 
- \left(\sum_{i=1}^n \sum_{j=1}^{N_i}  \int_0^{t_{ij}}\exp(n_{ij}(s)) \boldsymbol \Phi (s) \boldsymbol \Phi^{\top} (s)  ds + \boldsymbol A_{\zeta} \right),\\
&\boldsymbol{F}_{\boldsymbol \beta b_i}  = \boldsymbol{F}^{\top}_{b_i \boldsymbol \beta} && = - \sum_{j=1}^{N_i} \mathbf x_{ij} \mathbf u_{ij}^{\top} \int_0^{t_{ij}} \exp(\eta_{ij}(s)) ds,\\
&\boldsymbol{F}_{\boldsymbol \alpha  b_i}  = \boldsymbol{F}^{\top}_{b_i \boldsymbol \alpha} && = - \sum_{j=1}^{N_i}  \mathbf u_{ij}^{\top} \int_0^{t_{ij}} \exp(\eta_{ij}(s)) \boldsymbol \Phi(s) ds, \\
&\boldsymbol{F}_{b_i b_i}  && = - \left(\sum_{j=1}^{N_i} \mathbf u_{ij} \mathbf u_{ij}^{\top} \int_0^{t_{ij}} \exp(\eta_{ij}(s)) ds
+ \boldsymbol Q^{-1} \right),\\
&\boldsymbol{F}_{b_i b_j}  = \boldsymbol{F}_{b_j b_i} && = 
\boldsymbol 0.
\end{alignedat}
$
\end{center}

%\textcolor{red}{Bitte √ºberpr√ºfen, ggf. stichpunktartig schreiben, was fehlt}

\subsubsection*{Update of random effects variance-covariance parameters}

If random effects are included, the random effects variance update $\boldsymbol Q^{[l]}$ in each iteration $l$ 
(indicated by the superscript $[l]$) of the fitting routine is computed as
\begin{align*}
    \hat{\boldsymbol{Q}}^{[l]} = \frac{1}{n} \sum_{i=1}^n\left(\hat{\boldsymbol V}_{b_ib_i}^{[l]} + \hat{\boldsymbol b}_i^{[l]} (\hat{\boldsymbol b}_i^{[l]})^{\top}\right). 
\end{align*}
Using the notation $\tilde{\boldsymbol\beta}$ as shortcut for $\tilde{\boldsymbol\beta}=(\boldsymbol\beta^{\top}, \boldsymbol \alpha^{\top})^{\top}$, posterior curvatures $V_{b_i b_i}$ are calculated as
\begin{align*}
    \boldsymbol V_{b_i b_i} =  \boldsymbol F_{b_i b_i}^{-1} + \boldsymbol F_{b_i b_i}^{-1} \boldsymbol F_{b_i \tilde{\boldsymbol \beta}} (\boldsymbol F_{\tilde{\boldsymbol \beta} \tilde{\boldsymbol \beta}}
    - \sum_{i=1}^n  \boldsymbol F_{\tilde{\boldsymbol \beta} b_i}
    \boldsymbol F_{b_i b_i}^{-1} \boldsymbol F_{b_i \tilde{\boldsymbol \beta}})^{-1} \boldsymbol F_{\tilde{\boldsymbol \beta} b_i} \boldsymbol F_{b_i b_i}^{-1}\,.  
\end{align*}
This is based upon the standard maximum likelihood result that $\hat{\boldsymbol \delta} \stackrel{a}{\sim}  N(\boldsymbol \delta, \boldsymbol F(\hat{\boldsymbol \delta})^{-1})$ which is justified if cluster sizes are large enough and from inverting 
partitioned matrices \citep[see][e.g.]{Fahrmeir2001}.   

\subsubsection*{{Starting values}}
 
Regarding the starting values, we set the parameters $\hat{\boldsymbol{\beta}}^{(0)}$  and $\hat{\boldsymbol{b}}^{(0)}$ to zero while holding penalty parameter $\xi$ fixed. The starting vector  $\hat{\boldsymbol{\alpha}}^{(0)}$ includes all zeros except of the first entry, which is the first spline coefficient of the baseline hazard resembling the intercept value. It is set to the ML estimate of a baseline hazard that is constant over time and across individuals. The starting values for $\hat{\boldsymbol{\theta}}^{(0)}$ are chosen such that the starting covariance matrix for the random effects is diagonal with values $0.1$ as diagonal elements. The penalty parameter {$\zeta_0$} is firstly set to a value that shrinks $\hat{\boldsymbol{\alpha}}^{(0)}$ to zero. The same is done for the penalty parameters $\zeta_1,\ldots,\zeta_m$ if, in addition, $m$ time-varying coefficients are included.

%%%%%%%%%%%%%%%%%%%%%%%%%%%%%%%%%%%%%%%%%%%%%%%%%%%%%%%%%%%%%%%%%%%%%%%%%%%%%%%%%%%%%%
%%%%%%%%%%%%%%%%%%%%%%%%%%%%%%%%%%%%%%%%%%%%%%%%%%%%%%%%%%%%%%%%%%%%%%%%%%%%%%%%%%%%%%

\section{Simulation study}

% gemerelles vorgehen bei unserer simulation mit ändernden kovariablen: max zufällig 9 splits, 10 werte kovariable
% unabh. davon event times und censoring times simuliert
% ziehem  zufallszahl  , in -log eingesetzt (wir wollen auf ebene des baseline hazard kommen)
% klein lambda wird aufintegriert bis simulierte zahl erreich , das ist dann zeit, je nachdem ob diese größer oder kleiner censored zeit , ist es dann event oder censored, 

We conduct a simulation study to evaluate the performance of our implementation and compare it with predominant survival packages. For some of the scenarios, to our knowledge no comparable package that perform a LASSO regularisation exists  and we hence include also unregularised approaches to compare the estimated coefficients and baseline hazard.   

\subsection{Model setup and data generating process}

The underlying model used in both simulation studies is 
\begin{align}
    \lambda_{ij}(t| \mathbf{x}_{ij}, {b}_i) & = \lambda_0(t)  \exp(\eta_{ij}), \qquad i = 1,\dots, n, \quad j = 1, \dots, N_i\,, \\
    \textrm{with} \qquad \quad \eta_{ij}(t) &=  \mathbf x_{ij}^{\prime}\boldsymbol \beta + b_i\,, \nonumber   
\end{align}
although some simulation scenarios use versions of $\eta_{ij}$ without random effects $b_i$. Some covariates in $\mathbf{x}_{ij}$ change at several time points. The baseline hazard and the effects $\boldsymbol \beta$  take on the following forms:
\begin{align*}
  \lambda_0(t) &= 15 \cdot f_{\chi^2}(t) + 0.15,  \\    
   \beta_1 &= 0.6,\\ 
   \beta_2 &=-0.7,\\ 
   \beta_3 &= 0.4,\\
   \beta_4 &= -0.8, \\
   \beta_5 &= \ldots = \beta_{20} = 0,
\end{align*}
%
%\vskip 6pt
%\begin{tabular}{llll}
% \multicolumn{3}{l}{ $\lambda_0(t) = 15 \cdot f_{\chi^2}(t) + 0.15$,}  &  \\    
% $\beta_1 = 0.6$, &  \quad $\beta_2 =-0.7$, & \quad $\beta_3 = 0.4$, &  \quad $\beta_4 = -0.8$,  \\
%  \multicolumn{2}{l}{$\beta_5 = \ldots = \beta_{20} = 0$,} & &
%\end{tabular}
%\vskip 6pt
where $f_{\chi^2}$ is the density of a $\chi^2$-distribution $\chi^2(df, \delta)$ with degrees of freedom $df=14$ and non-centrality parameter $\delta = 2$. 
The effects $\beta_5$ to $\beta_{20}$  add noise to investigate the performance of \texttt{coxlasso} regarding variable selection. 
The covariates are drawn independently from a uniform distribution, i.e.\ $x_{ijk}\sim U(0,1), k = 1, \dots, 20$. 
The number of observations is $n = 500$  and in scenarios that include random effects, these observations are evenly clustered into 50 groups such that each group comprises $N_i=10$ observations.  The random effects are normally distributed with $b_i \sim N(0, 1)$. 
Censoring times are drawn from a uniform distribution on the $[0;10]$ interval. 

To compare the performance of \texttt{coxFL} and \texttt{coxlasso} with more established routines, we evaluate the estimates of baseline hazard, covariate effects, and random effects variance separately. Averaging over 100 simulation runs, mean squared errors are applied as follows:  
\begin{align*}
&\textrm{MSE}_{\lambda_0} := \sum_{t=1}^T \omega_t(\lambda_0 - \hat{\lambda}_0)^2, \qquad \textrm{MSE}_{\beta_k} &:= \sum_{k=1}^1 \sum_{t=1}^T \omega_t(\beta_k - \hat{\beta_k})^2, \qquad \textrm{MSE}_{\sigma_b} := (\sigma_b - \hat{\sigma}_b)^2, \\
&\textrm{with} \quad \omega_t = \frac{\Lambda_0(T) - \Lambda_0(t)}{\Lambda_0(T)}\,,
\end{align*}
where $\omega_t$ are weights that are based on the cumulative baseline hazard $\Lambda_0(\cdot)$. We also compute the proportions of correctly selected variables when using \texttt{coxlasso}. % and record non-convergent simulation runs and the computational time.

%%%%%%%%%%%%%%%%%%%%%%%%%%%%%%%%%%%%%%%%%%%%%%%%%%%%%%%%%%%%%%%%%%%%%%%%%%%%%%%%%%%%%%
\subsection{Simulation scenarios for \texttt{coxlasso}}

The effectiveness of our approach is assessed in four different scenarios: 
\begin{flalign*}
 & \textbf{Scenario 1}  \qquad 
 \eta_{ij}(t) =   \sum_{k=1}^{20} x_{ijk}\beta_k,  &     \\
  & \textbf{Scenario 2}  \qquad 
     \eta_{ij}(t) =  \sum_{k=1}^{20} x_{ijk}\beta_k + b_i ,&     \\
   % 3 vary
  & \textbf{Scenario 3}  \qquad 
     \eta_{ij}(t) = \sum_{k = 1}^4 x^{(t)}_{ijk} \beta_k   + \sum_{k=5}^{20} x^{(t)}_{ijk}\beta_k,&  \\   
 % 4 vary
 & \textbf{Scenario 4}  \qquad 
     \eta_{ij}(t) =  \sum_{k = 1}^4 x^{(t)}_{ijk} \beta_k  + \sum_{k=5}^{20} x^{(t)}_{ijk}\beta_k + b_i,     
\end{flalign*}
where the superscript $(t)$ indicates that these covariates change several times over the time period (up to a maximum of 10 changes per observation). 
That is, we start with a simple Scenario~1, where covariates are only time-constant and no random effects are included. We then first add either random effects (Scenario~2) or time-varying covariates (Scenario~3), or finally both (Scenario~4). 

The results of all scenarios are compared to popular existing packages that apply to the specific scenarios. For Scenario~4, no existing package can be applied. Table \ref{tab:benefits} gives an overview of popular packages for fitting a Cox model and demonstrates the differences to \texttt{coxlasso}. While \texttt{coxph} and \textit{coxme} do not perform variable selection, other packages that can include a classic lasso penalty in a Cox model are \texttt{penalized} and \texttt{glmnet}. Both of these do not estimate a smooth baseline hazard nor do they include random effects. Principally, time-varying covariates are possible in \texttt{penalized}, but then often huge internal matrices are created when the number of observations is large, using large amounts of working memory in these cases, sometimes even leading to a program crash.    
None of the mentioned existing packages for estimating Cox models is based on the full likelihood.

\begin{table}[ht]
\centering
\caption{Comparison of \texttt{coxlasso} with other packages for regularized Cox models}\label{tab:benefits}
\vspace{6pt}
\begin{tabular}{p{1.8cm}p{1.5cm}p{2.2cm}p{1.7cm}p{1.7cm}p{1.6cm}p{1.6cm}}
\hline 
& LASSO & baseline hazard & time- \linebreak varying \linebreak coefficients & time- \linebreak varying \linebreak covariates & random \linebreak effects  & full \linebreak likelihood \\
\hline
\texttt{coxph} & {no} & Nelson-Aalen & user$^{\ast}$  &  {yes} &  {yes} & {no} \\
\texttt{coxme} & {no} & n.a. 
&  user$^{\ast}$ & {yes} & {yes} & {no}  \\
\texttt{penalized} & {yes} & Breslow & {no} &not for n$\uparrow^{\dag}$ & {no} & {no} \\
\texttt{glmnet} & {yes} & Breslow with \texttt{hdnom} & {no} &{no} &  {no} &  {no}  \\
\texttt{coxlasso} & {yes} & B-splines & {yes} & {yes} & {yes} & {yes}\\
\hline
%\multicolumn{7}{l}{\small{Note: Abbreviations "tv coef", "tv coef"}}
\end{tabular}
\raggedright $^{\ast}$ {\small ``user'' means that the functional form of the time-varying coefficient must be specified by the user.} \\
\raggedright $^{\dag}$ {\small ``not for n$\uparrow$'' means that the package has troubles when dealing with a large number of observations. }\\
\end{table}

%%%%%%%%%%%%%%%%%%%%%%%%%%%%%%%%%%%%%%%%%%%%%%%%%%%%%%%%%%%%%%%%%%%%%%%%%%%%%%%%%%%%%%
%%%%%%%%%%%%%%%%%%%%%%%%%%%%%%%%%%%%%%%%%%%%%%%%%%%%%%%%%%%%%%%%%%%%%%%%%%%%%%%%%%%%%%

\subsection{Simulation results}

\begin{figure}
    \centering
  \includegraphics[scale = 0.8]{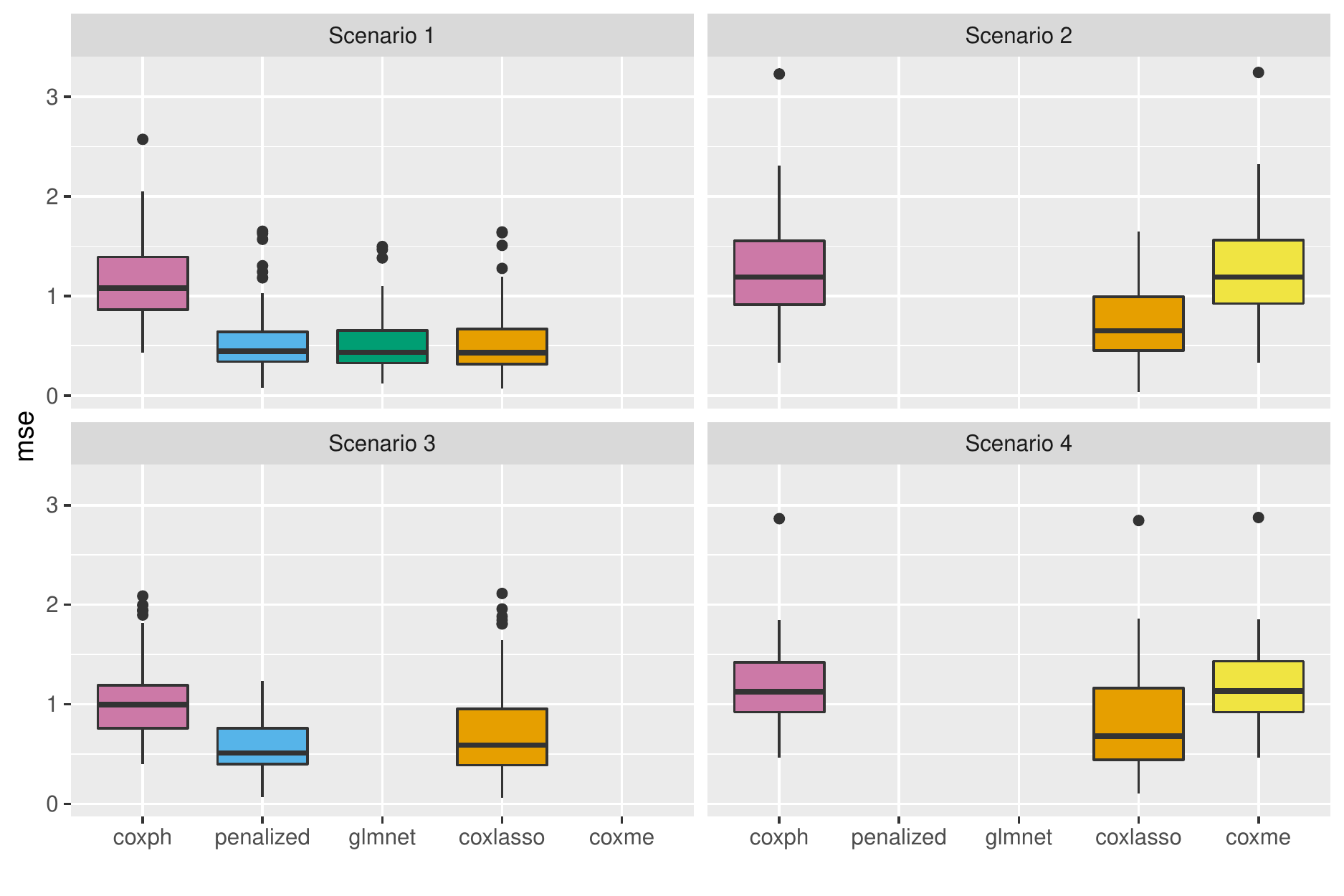}
    \caption{Boxplots of linear coefficients' MSE for different \textit{R} packages. 
     Note that \texttt{coxph} and \texttt{coxme} cannot perform variable selection and that \texttt{coxme} was specifically designed for frailty models and is thus not used for Scenario~1 and Scenario~3.}
    \label{fig:boxplot_linear}
\end{figure}

\begin{figure}
    \centering
    \includegraphics[scale = 0.8]{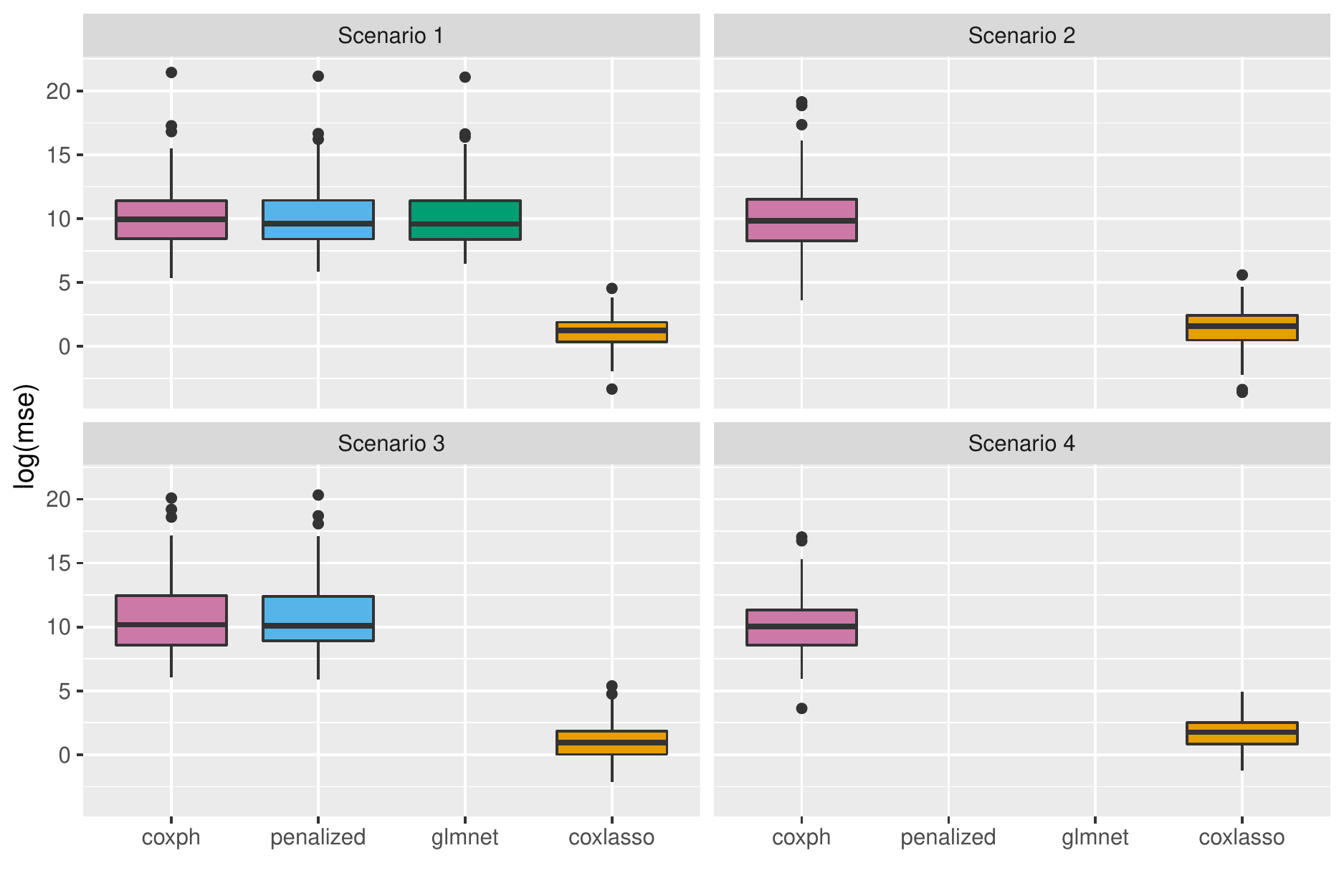}
    \caption{Boxplots of log(MSE) for estimates of the baseline hazard. As \texttt{coxme} does not allow to extract the baseline hazard it is not shown here.}
    \label{fig:boxplot_baseline}
\end{figure}

Simulation results for all results are shown in Figures~\ref{fig:boxplot_linear} and \ref{fig:boxplot_baseline} and Table~\ref{tab:sim_res}.
For the linear coefficients, the simulation exercise confirms that our functions \texttt{coxlasso} performs comparable to the other established \texttt{R} routines in terms of the MSE. In Scenario~3, \texttt{penalized} performs slightly better but it does not allow to include random effects, time-varying coefficients nor time-varying covariates. Therefore, in Scenario~4 we can compare \texttt{coxlasso} only to \texttt{coxph} and \texttt{coxme} which, however, do not perform variable selection. \texttt{coxlasso} outperforms the other packages in terms of the MSE even though its variance somewhat increases. The lower MSE supports our suspicion that the full likelihood is beneficial when covariates change frequently.   

In terms of the baseline hazard, \texttt{coxlasso}  substantially outperforms the other packages as it estimates a smooth baseline hazard and thus captures it better than the Breslow estimator used in the other functions. The package \texttt{coxme} does not allow to extract the baseline hazard and \texttt{glmnet} requires \texttt{hdnom} to report the baseline hazard. The high MSE for the packages we used for comparison results from their step function approach on the cumulative baseline hazard level, whose first derivative then needs to be approximated. 
%\textcolor{red}{here more info, breslow, what do we plot? F? of baseline?} 

To evaluate the variable selection performance, Table~\ref{tab:sim_res} also reports the true positive rate (TPR) and false discovery rate (FDR). While \texttt{coxphh} and \texttt{coxme} do not perform variable selection, i.e.\ they always have a TPR of 1, \texttt{coxlasso} performs well and comparable to \texttt{penalized} and \texttt{glmnet}. 

Note that we focused in the simulations on estimating the effects of changing covariates and the baseline hazard and did not explicitly include time varying coefficients. However, the underlying mechanism for the baseline hazard and time varying coefficients is the same as the baseline hazard can be understood as a covariate of value one with time varying effects. 
Overall, the simulation study showed that \texttt{coxlasso} works well and gives robust estimates across all four considered scenarios. 

\begin{table}[ht]
\centering
\caption{Simulation results} \label{tab:sim_res}
\begin{tabular}{llrrrrr}
  \hline
scenario & criteria & coxph & coxme & penalized & glmnet & coxlasso \\ 
  \hline
1 & mean mse baseline & 21548496.37 & - & 16148321.14 & 15033469.85 & 7.13 \\ 
  1 & mean mse lin. coef & 1.14 & - & 0.52 & 0.51 & 0.51 \\ 
  1 & TPR selection & - & - & 0.92 & 0.9 & 0.98 \\ 
  1 & FDR selection & - & - & 0.56 & 0.49 & 0.72 \\ 
\hline 
  2 & mean mse baseline & 4373107.9 & - & - & - & 14.02 \\ 
  2 & mean mse lin. coef & 1.25 & 1.26 & - & - & 0.72 \\ 
  2 & mean mse re & 11.16 & 11.2 & - & - & 10.94 \\ 
  2 & mean mse re var & 0.05 & 0.07 & - & - & 0.02 \\ 
  2 & TPR selection & - & - & - & - & 0.76 \\ 
  2 & FDR selection & - & - & - & - & 0.40 \\ 
 \hline  
  3 & mean mse baseline & 9664549.09 & - & 9840842.99 & - & 8.01 \\ 
  3 & mean mse lin. coef & 1.03 & - & 0.56 & - & 0.72 \\ 
  3 & TPR selection & - & - & 0.94 & - & 0.98 \\ 
  3 & FDR selection & - & - & 0.94 & - & 0.73 \\ 
 \hline
  4 & mean mse baseline & 661283.33 & - & - & - & 11.03 \\ 
  4 & mean mse lin. coef & 1.17 & - & - & - & 0.82 \\ 
  4 & TPR selection & - & - & - & - & 0.97 \\ 
  4 & FDR selection & - & - & - & - & 0.73 \\ 
   \hline
\end{tabular}
\end{table}

%%%%%%%%%%%%%%%%%%%%%%%%%%%%%%%%%%%%%%%%%%%%%%%%%%%%%%%%%%%%%%%%%%%%%%%%%%%%%%%%%%%%%%
%%%%%%%%%%%%%%%%%%%%%%%%%%%%%%%%%%%%%%%%%%%%%%%%%%%%%%%%%%%%%%%%%%%%%%%%%%%%%%%%%%%%%%

\section{Application cases}

We apply \texttt{coxlasso} to two application cases: The Duration of exclusive breast feeding in Indonesia and to genetic data of non-small cell lung cancer (NSCLC). For the first case, we perform variable selection within a set of several covariates. For the latter case, we explicitly decide that a clinical covariate (smoking) should be part of the model and that its effect can be varying over time. Variable selection is then performed within a large set of potential covariates including genetic information. 

\subsection{Application 1: Duration of exclusive breastfeeding in Indonesia}

In our first application, we analyze the determinants of the duration of exclusive breastfeeding in Indonesia using data from the Indonesian Family Life Survey East 2012 \citep{IFLS.2012}. Due to the positive impact of breastfeeding on children's health \citep[see, e.g.,][for a review]{Dieterich.2013}, understanding its drivers helps designing effective awareness campaigns and health policies.

As response variable we use the duration of exclusive breast feeding (in months) whose end point is the event when the baby is fed water or other food for the first time. Exclusive breast feeding duration is related to covariates characterizing the household, the mother, health care demand, and health care supply. The full set of potential covariates is drawn from \citet{LoBue.2018}, who kindly provided us with their data cleaning code to replicate their sample. Our example differs from their analysis in that we perform a time-to-event analysis and also include children that are still being breastfeed, i.e.\ we account for right-censoring. More precisely, while they analyzed the drivers for the optimal duration of exclusive breastfeeding of six months in a linear probability model, we look at the crude duration rather than on a specific length. % and 
The final data set comprises 1,193 children and 20 covariates, including household characteristics, mothers' education and age, and health care supply on the community level. % Mothers are modelled as random effects.  
\begin{figure}\centering
\begin{subfigure}[t]{0.5\textwidth}
 \includegraphics[scale = 0.5]{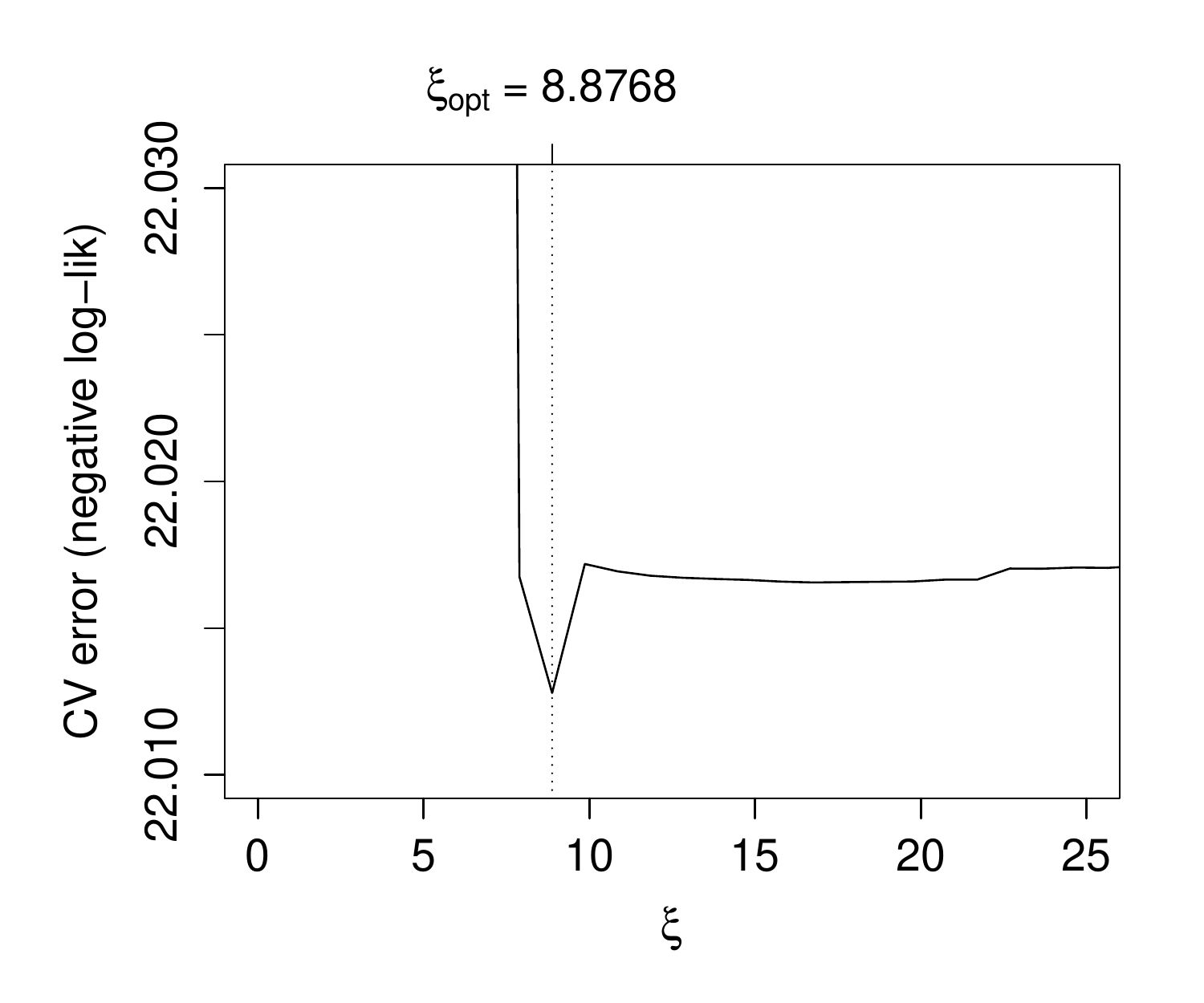}
\subcaption{Cross validation results} \label{fig:cv_breast}
\end{subfigure}\\
\begin{subfigure}[t]{0.5\textwidth}
   \includegraphics[scale = 0.5]{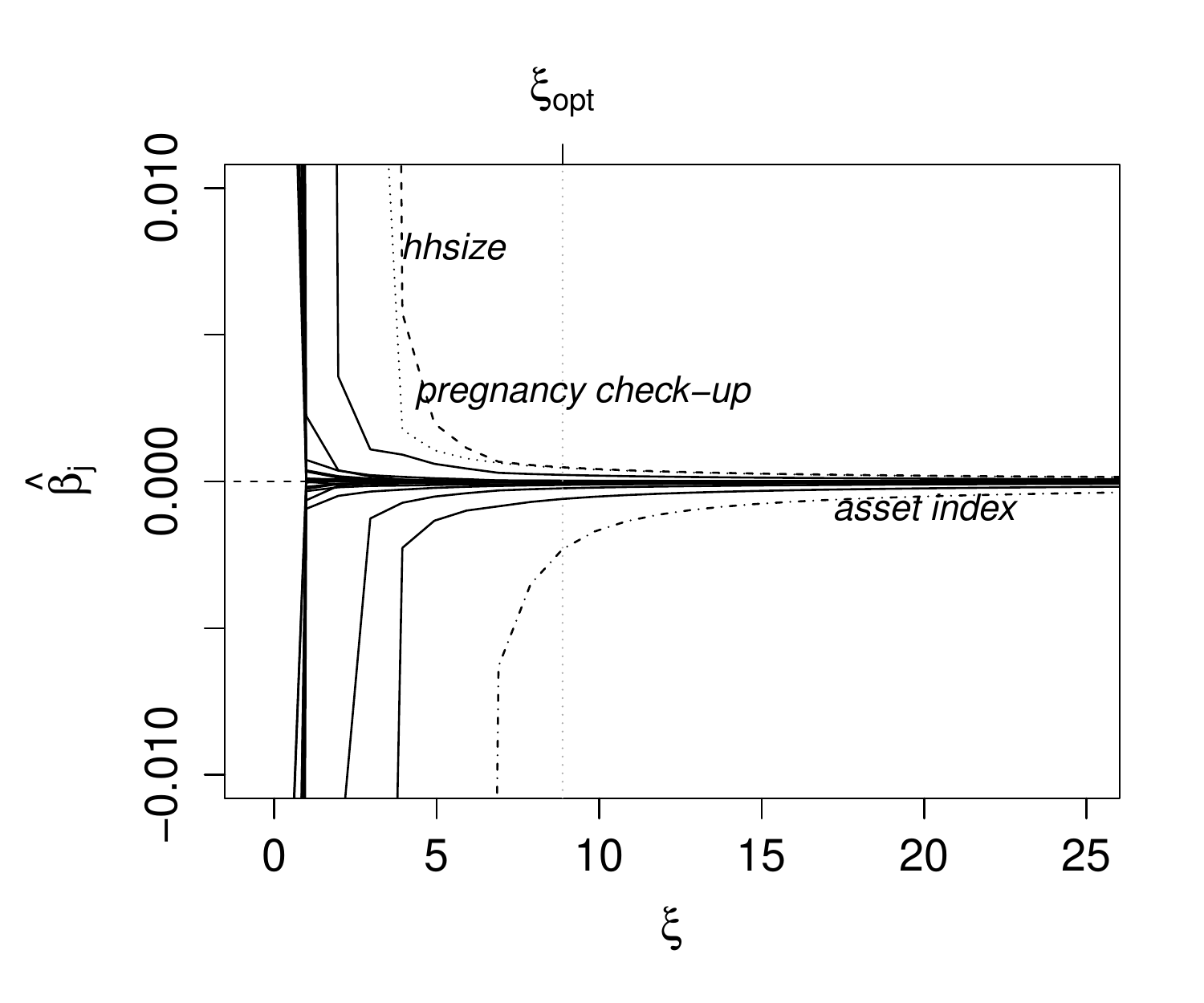}
\subcaption{Coefficient paths vs.\ LASSO penalty parameter $\xi$; optimal value for $\xi$ is shown by dotted vertical line}  \label{fig:coef_path}
\end{subfigure}\\
\begin{subfigure}[t]{0.5\textwidth}
 \includegraphics[scale = 0.5]{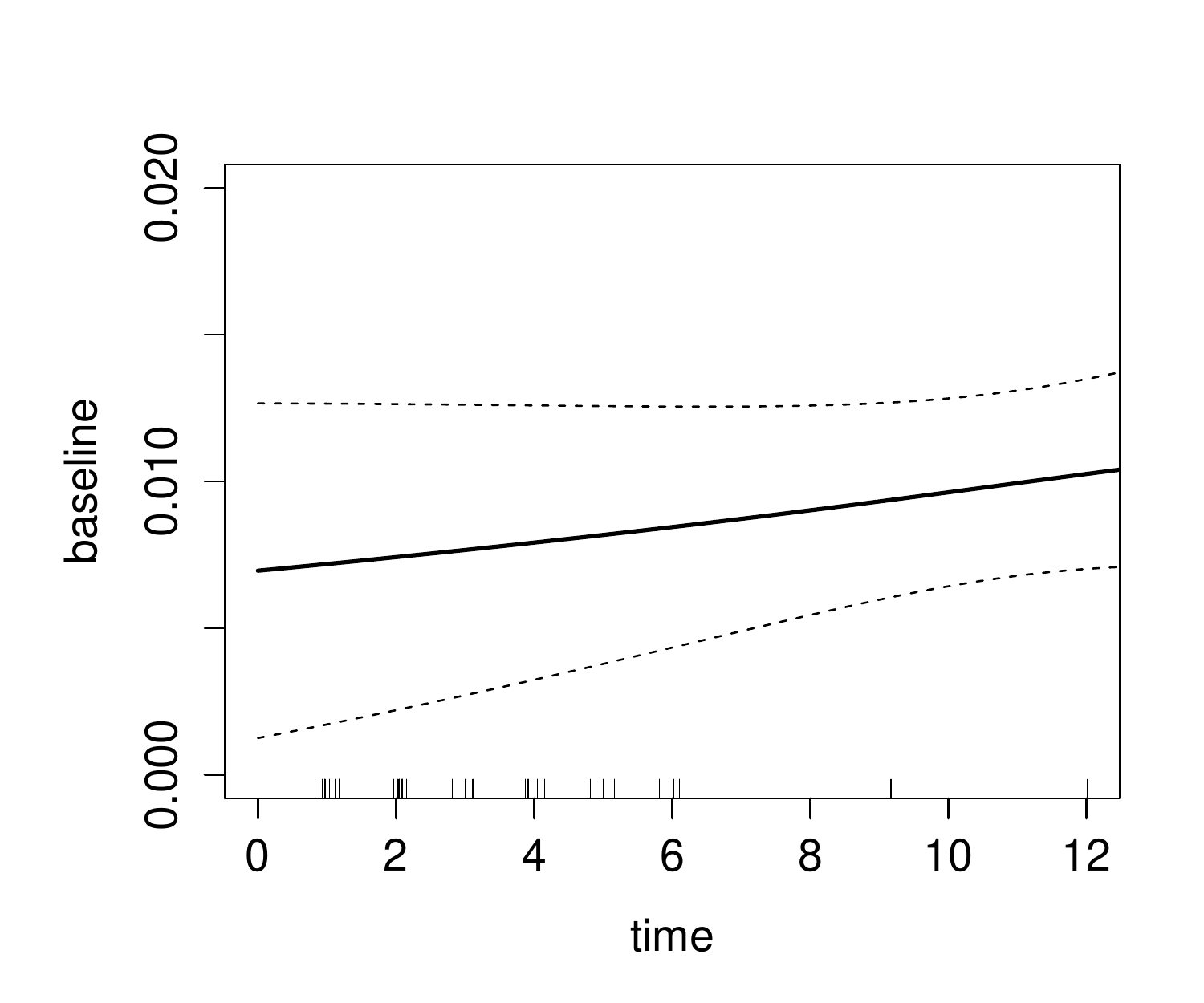}
\subcaption{Smooth estimate of baseline hazard} \label{fig:base_breast}
\end{subfigure}
\vskip 6pt
\caption{Results for the breastfeeding example. %Grey dashed lines in subfigure (a) indicate one standard deviation of the cross validation error. 
In subfigure (b) the dotted line belongs to the variable {\it pregnancy check-}up, the dashed line to {\it household size}, and the dotted-dashed line to the {\it asset index}. Black dashed lines in subfigure (c) display point-wise 95\% confidence intervals.}
\end{figure}

\begin{table}[ht]
\centering
\caption{Coefficients of selected determinants of exclusive breast feeding duration in Indonesia \label{tab:coef_indo}} 
\begin{tabular}{lc} 
  \hline
selected variable & coefficients \\ 
  \hline
birth order & -0.00005 \\ 
  hh size & 0.00008 \\ 
  asset index & -0.00068 \\ 
  pregnancy services & 0.00004 \\ 
  pregnancy checkup & 0.00056 \\ 
  \hline  
\multicolumn{2}{p{10cm}}{Note: Variables that were not selected include the gender of the child, dummy for short birth interval, mother's age at birth, mother's years of schooling, and whether the mother has a formal job, whether a grandmother lives in the household, and several health care variables.}
\end{tabular}
\end{table}

Table \ref{tab:coef_indo} shows the results for the estimated coefficients.
The LASSO approach selects the following set of predictors to be relevant: \textit{birth order}, \textit{household size}, \textit{asset index}, whether the mother had a \textit{pregnancy checkup} and how many \textit{pregnancy services} she used. The \textit{asset index} seems to be a strong predictor since it enters the model first, which is shown in the coefficient built-up paths in Figure~\ref{fig:coef_path}. A possible explanation for the negative sign of the \textit{asset index} might be that wealthier families are more likely to purchase store-bought baby  food, whereas poorer families rely more on inexpensive breast milk. Health care supply on the community level seems to be beneficial as both variables \textit{pregnancy services} and \textit{pregnancy check up}  are associated with a longer duration on exclusive breastfeeding.  
Figure~\ref{fig:cv_breast} shows the CV error against the penalty parameter $\xi$, which chooses a penalty of $\xi_{opt} = 8.88$. The shape of the curve suggests that the effect of overfitting increases very strongly with penalties smaller than $\xi_{opt}$ while the effect of underfitting, i.e.\ penalties larger than $\xi_{opt}$, is relatively flat. The standard deviation of the CV error is relatively large in relation to the shown image section and thus no intervals are printed here. Consequently, the chosen model is relatively sparse at the optimal amount of penalization as most coefficient paths begin after the vertical dashed line in Figure~\ref{fig:coef_path}, representing $\xi_{opt}$.        
The baseline hazard is fitted using six basis functions and is displayed in Figure~\ref{fig:base_breast}, looking rather linear.

%%%%%%%%%%%%%%%%%%%%%%%%%%%%%%%%%%%%%%%%%%%%%%%%%%%%%%%%%%%%%%%%%%%%%%%%%%%%%%%%%%%%%%
%%%%%%%%%%%%%%%%%%%%%%%%%%%%%%%%%%%%%%%%%%%%%%%%%%%%%%%%%%%%%%%%%%%%%%%%%%%%%%%%%%%%%%

\subsection{Application 2: Genomic biomarkers of lung cancer}

The second example deals with lung tumor samples and includes both, clinical and genetic data. The dataset was kindly provided by \citet{Madjar.2018}, who drew the raw data from the Gene Expression Omnibus \citep{Edgar.2002} and curated them manually, i.e.\ removing duplicates, non-tumorous samples and observations with missing survival times. 

The dataset we use consists of four non-small cell lung cancer (NSCLC) cohorts and comprises 635 observations. As potential covariates we include five clinical variables namely, \textit{age}, \textit{smoking}, \textit{stage}, \textit{gender}, and \textit{histology}. \textit{Smoking} is modeled as a time-varying coefficient, i.e.\ it is not prone to variable selection and we expect varying effects between former or current smokers and never smokers. \textit{Age} is metric, \textit{gender} is a dummy variable and \textit{stage} and \textit{histology} are coded as factor variables. The \textit{histology} categories include adenocarcinoma, squamous cell carcinoma, and other NSLCCs. Unfortunately, the numbers in the histological categories are quite unbalanced with only 34 observations in category~3. For illustration purposes, we excluded observations from the third category, which yielded ``better'' plots of the CV results, but affected final variable selection only mildly. 

\begin{figure}\centering
\begin{subfigure}[t]{0.5\textwidth}
 \includegraphics[scale = 0.5]{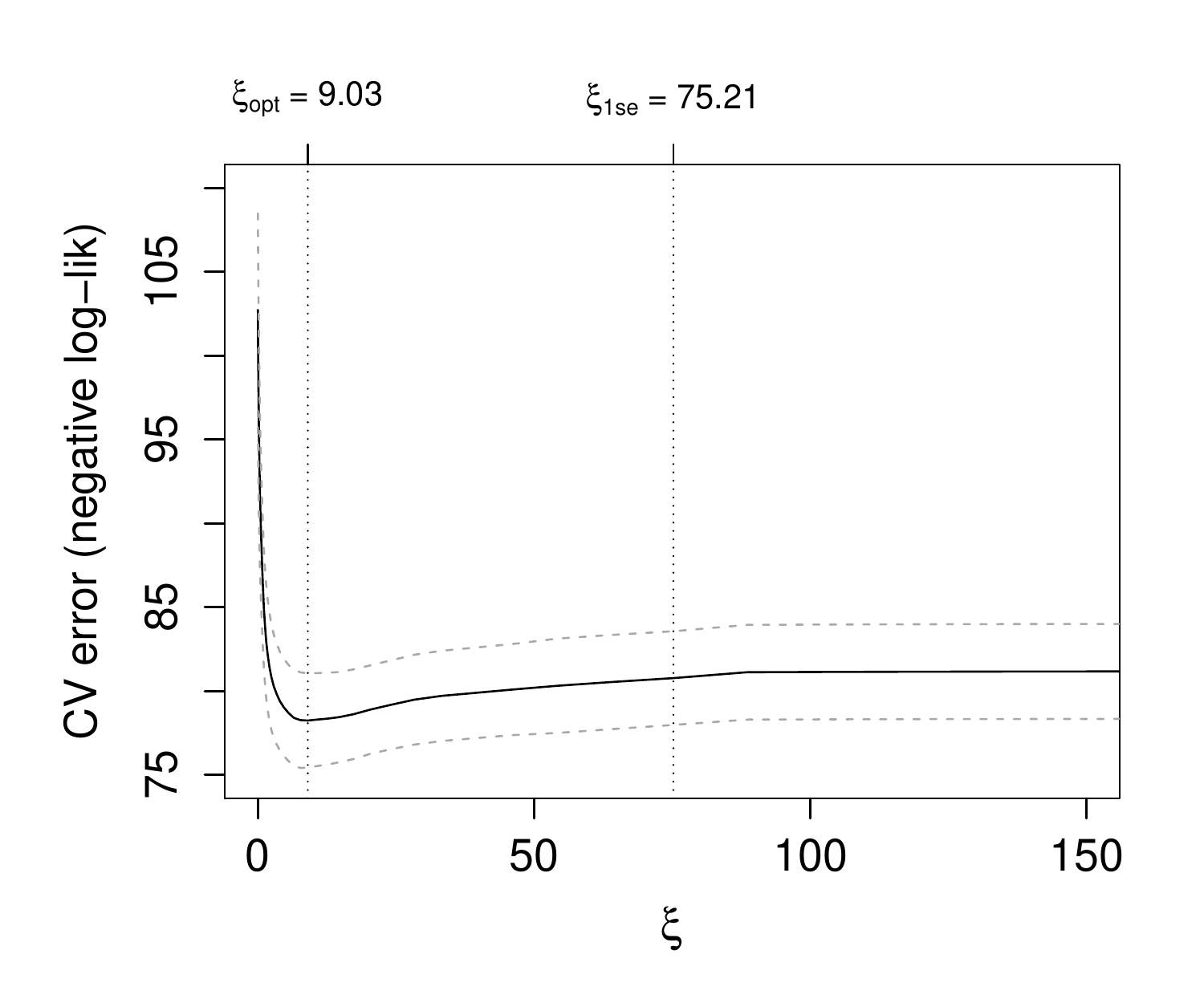}
\subcaption{Cross validation results} \label{fig:cv_genes}
\end{subfigure}%
\begin{subfigure}[t]{0.5\textwidth}
   \includegraphics[scale = 0.5]{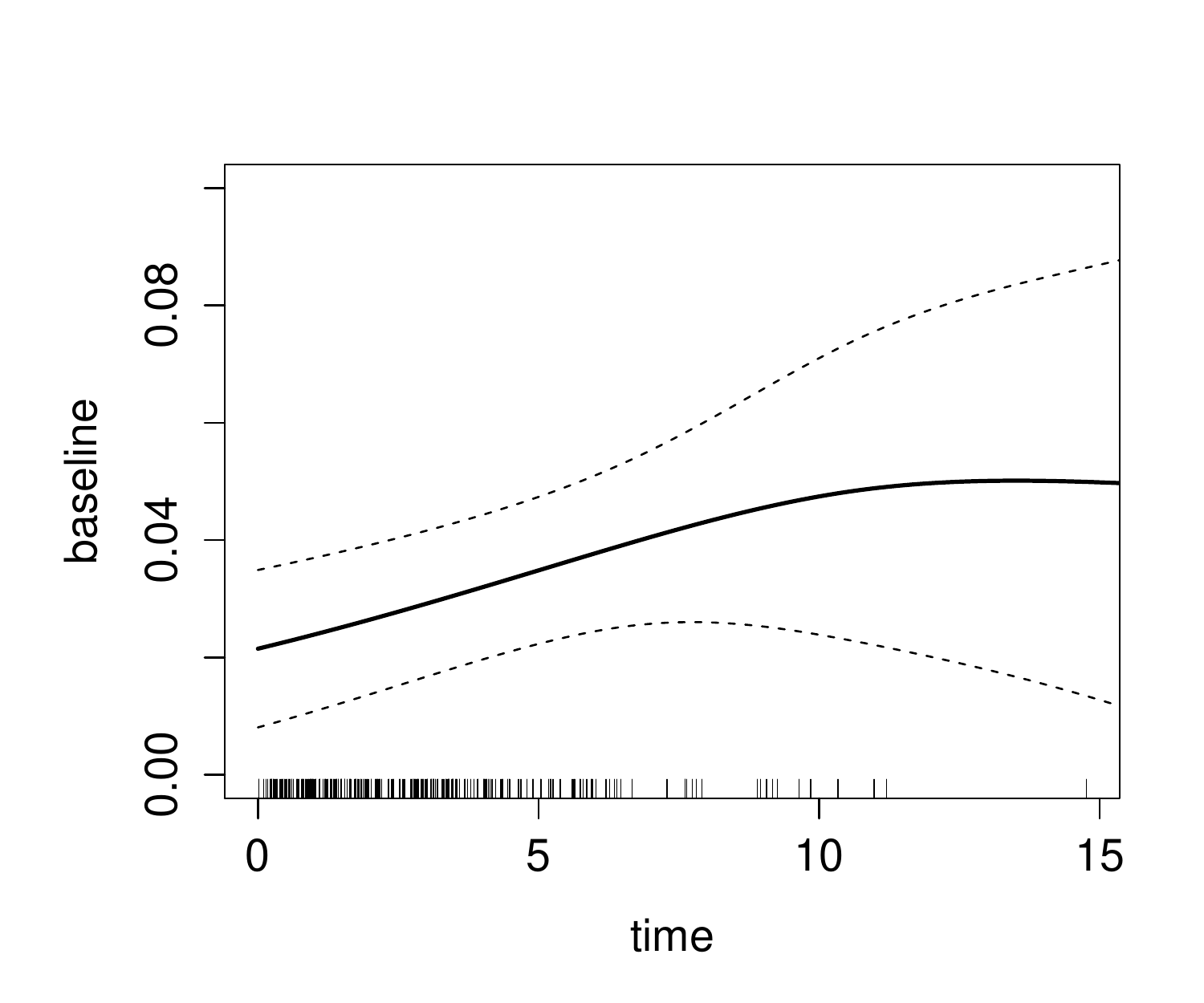}
\subcaption{Smooth estimate of baseline hazard}  \label{fig:base_genes}
\end{subfigure}\\
\begin{subfigure}[t]{0.5\textwidth}
 \includegraphics[scale = 0.5]{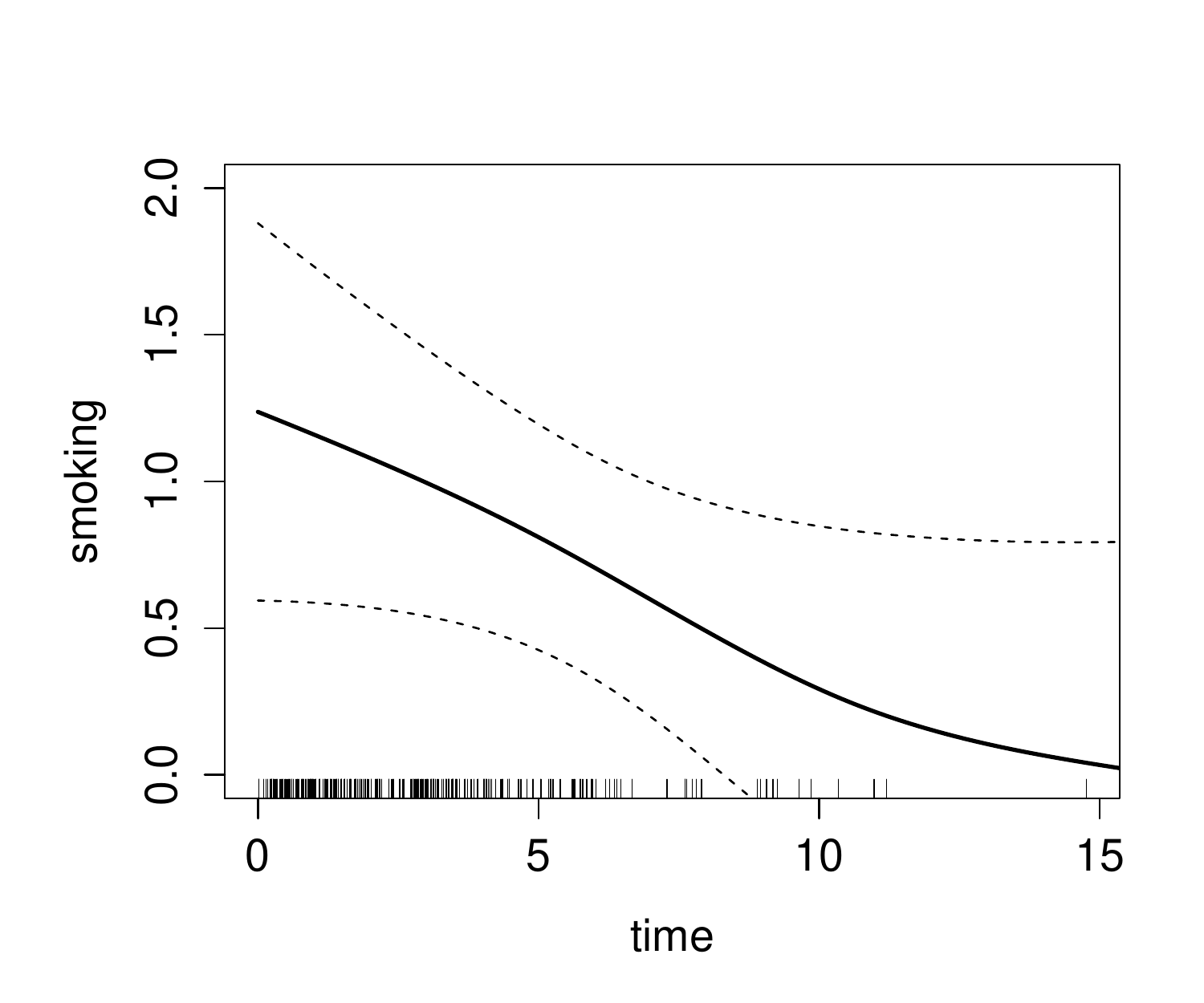}
\subcaption{Estimate of time-varying effect of smoking} \label{fig:vary_coef}
\end{subfigure}%
\begin{subfigure}[t]{0.5\textwidth}
 \includegraphics[scale = 0.5]{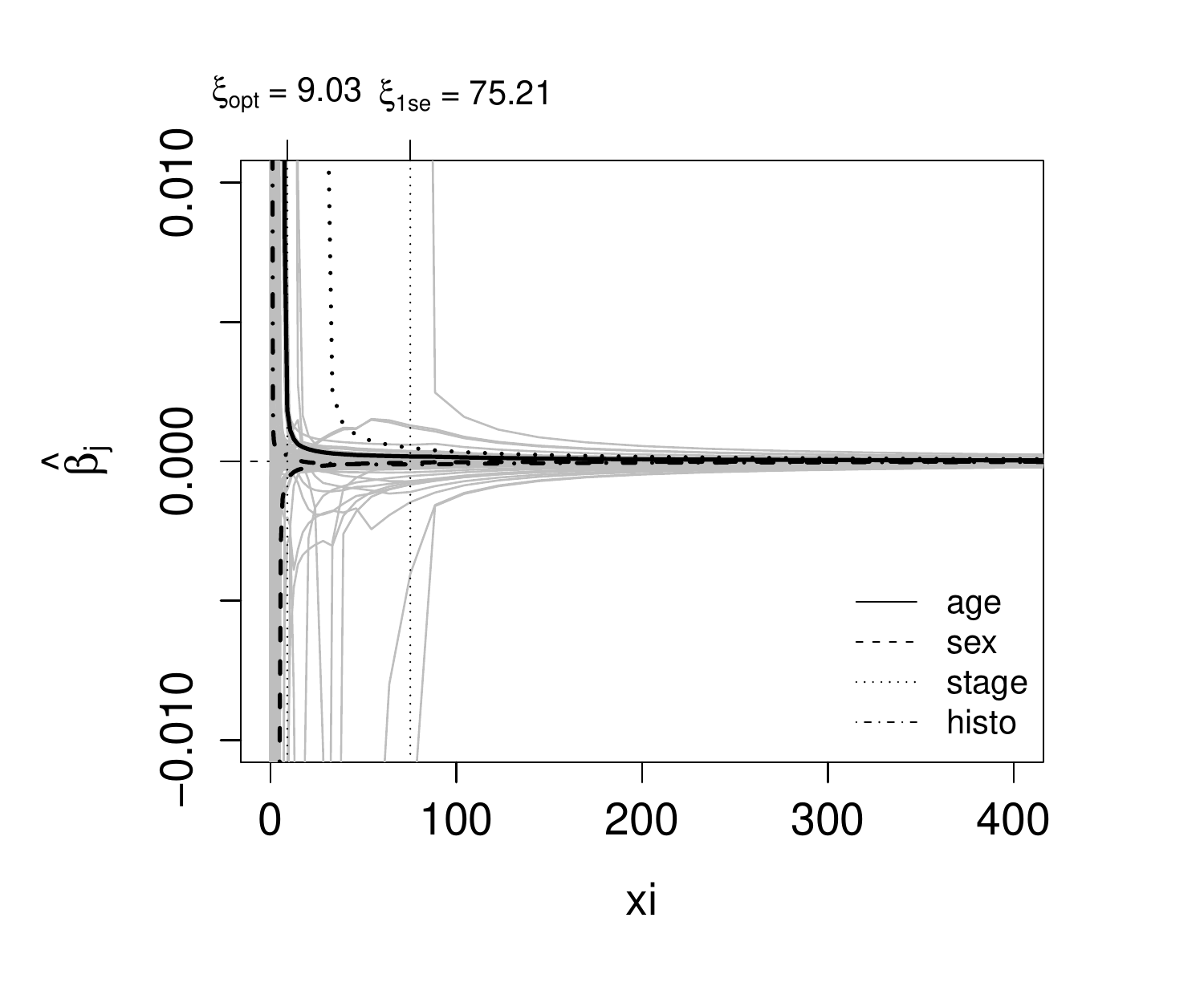}
\subcaption{Coefficient paths} \label{fig:coef_genes}
\end{subfigure}
\vskip 6pt
\caption{Results for the lung cancer data set. Grey dashed lines in subfigure (a) indicate one standard deviation of the CV error. Black dashed lines in subfigures (b) and (c) display point-wise 95\% confidence intervals. Vertical dotted lines in subfigure (d) indicate
penalty parameter values $\xi_{\text{opt}}$ (left) and $\xi_{\text{1se}}$ (right)}
\end{figure}

The CV plot in Figure~\ref{fig:cv_genes} shows nicely the trade off between overfitting and underfitting with the latter one having a milder effect on the CV error. The minimum yields a penalty strength of $\xi_{opt} = 9.03$. However, we decided to take the penalty $\xi_{1se} = 75.21$, whose CV error is within one CV standard error of $\xi_{opt}$, to obtain a sparser model. The model selects 43 covariates including \textit{age}, \textit{histology}, and \textit{stage}, where the latter might be the most important predictor out of these three as it is selected first, see Figure~\ref{fig:coef_genes}. This figure further shows that different genes have quite different effects on the hazard with some of them being even a stronger predictor than the clinical variables \textit{age}, \textit{sex}, \textit{stage}, and \textit{histology}.  

For \textit{smoking}, the estimated time-varying effect is presented in Figure~\ref{fig:vary_coef}. For earlier time points, being a former or current smoker has a positive effect on the hazard which is decreasing over time. This means the longer the patient survived after being diagnosed with NSCLC, the smaller the influence that the patient used to be a smoker -- assuming that patients stop smoking after the diagnosis. The estimate of the baseline hazard in Figure~\ref{fig:base_genes} is increasing but flattens for later time points. However, the rug plots at the bottom of both figures reveal that there are only few observations with late event points such that these later estimates are rather uncertain as indicated by the widening credibility intervals for these later time points.

%%%%%%%%%%%%%%%%%%%%%%%%%%%%%%%%%%%%%%%%%%%%%%%%%%%%%%%%%%%%%%%%%%%%%%%%%%%%%%%%%%%%%%
%%%%%%%%%%%%%%%%%%%%%%%%%%%%%%%%%%%%%%%%%%%%%%%%%%%%%%%%%%%%%%%%%%%%%%%%%%%%%%%%%%%%%%

\section{Concluding remarks}

This work proposes a flexible regularized Cox frailty model that is based on the full likelihood. Using the framework of the full likelihood has the advantage that we can directly estimate a smooth baseline hazard via P-splines and include both time-varying covariates and time-varying effects. Smoothing is carried out using a mixed model representation of the spline coefficients. Covariates are regularized using a lasso penalty with adaptive weights and categorical variables are penalized using a group lasso. This combination of flexible modeling with different effects and a lasso penalty in a full likelihood framework is implemented in the \texttt{R} function \texttt{coxlasso}. In a simulation study, our implementation shows good performance in terms of estimating both the regression coefficients and the baseline hazard, as well as in terms of variable selection. Since existing packages typically estimate the baseline hazard using a step function approach, \texttt{coxlasso} clearly outperforms them in this regard. We present two application cases using breastfeeding data from Indonesia and genetic data on small cell lung cancer. We particularly suggest using  \texttt{coxlasso} in rather high-dimensional settings in which (1) there are time-varying covariates and/or (2) there is a cluster structure and/or (3) when there are some important variables of interest that might have a time-varying effect that should be part of the model and a larger amount of covariates that are subject to penalization and variable selection.

% Fragen: time varying coefficients nicht simuliert

%%%%%%%%%%%%%%%%%%%%%%%%%%%%%%%%%%%%%%%%%%%%%%%%%%%%%%%%%%%%%%%%%%%%%%%%%%%%%%%%%%%%%%
%%%%%%%%%%%%%%%%%%%%%%%%%%%%%%%%%%%%%%%%%%%%%%%%%%%%%%%%%%%%%%%%%%%%%%%%%%%%%%%%%%%%%%

\bibliographystyle{kluwer}
\interlinepenalty=10000
\bibliography{diss_lit}

%%%%%%%%%%%%%%%%%%%%%%%%%%%%%%%%%%%%%%%%%%%%%%%%%%%%%%%%%%%%%%%%%%%%%%%%%%%%%%%%%%%%%%
%%%%%%%%%%%%%%%%%%%%%%%%%%%%%%%%%%%%%%%%%%%%%%%%%%%%%%%%%%%%%%%%%%%%%%%%%%%%%%%%%%%%%%

\end{document}